\documentclass[showpacs,prl,twocolumn,aps,superscriptaddress,preprintnumbers,letterpaper]{revtex4}
\usepackage{amsmath,amssymb}
\usepackage{epsfig}
\usepackage{graphicx}
\usepackage{amsmath}
\usepackage{amsfonts}
\usepackage{epstopdf}
\newcommand{\be}{\begin{equation}}
\newcommand{\ee}{\end{equation}}
\newcommand{\bear}{\begin{eqnarray}}
\newcommand{\eear}{\end{eqnarray}}
\newcommand{\ba}{\begin{array}}
\newcommand{\ea}{\end{array}}

\begin{document}

\title{Anomaly-induced Quadrupole Moment\\ 
of the Neutron in Magnetic Field}

\author{Dmitri E. Kharzeev}
\affiliation{Department of Physics and Astronomy, Stony Brook University, Stony Brook, New York 11794-3800, USA}
\affiliation{Department of Physics,
Brookhaven National Laboratory, Upton, New York 11973-5000, USA}

\author{Ho-Ung Yee}
\affiliation{Department of Physics and Astronomy, Stony Brook University, Stony Brook, New York 11794-3800, USA}

\author{Ismail Zahed}
\affiliation{Department of Physics and Astronomy, Stony Brook University, Stony Brook, New York 11794-3800, USA}

\date{\today}

\begin{abstract}
The neutrons cannot possess a quadrupole moment in the vacuum. Nevertheless, we show that in the presence of an external magnetic field the neutrons acquire a new type of quadrupole moment $Q^{ij}= \chi\,\sigma^i B^j$ involving the components of spin and magnetic field. This ``chiral magnetic" quadrupole moment arises from the interplay of the chiral anomaly and the magnetic field; we estimate its value for the neutron in the static limit, and find $\chi \simeq 1.35\cdot10^{-2}\,{\rm fm}^4$. The detection of 
 the quadrupole moment of the neutron would provide a novel test of the role of the chiral anomaly in low-energy QCD and can be possible in the presence of both magnetic and inhomogeneous electric fields. The quadrupole moment of the neutron may affect e.g. the properties of neutron stars and magnetars. 
%

\end{abstract}

\maketitle

\setcounter{footnote}{0}


{\bf 1.\,}
Quantum anomalies are known to play an important role in QCD. 
For example, the Abelian chiral anomaly \cite{Adler:1969gk,Bell:1969ts} accounts for the $\pi^0\rightarrow\gamma\gamma$ decay
and forces chiral Skyrmions to be baryons. 
In this short letter we will explore the effect of the chiral anomaly on the deformation of the neutron in an external magnetic field. 

The electric dipole moment of the neutron has attracted a lot of attention as it allows to put a strict bound on the amount of {\cal CP} violation in QCD and beyond, see \cite{Baker:2006ts} for a recent result. 
The electric quadrupole moment defined as the expectation value of the operator $\hat{Q}_E^{ij} = \sum e (3 x^i x^j - \delta^{ij} x^2)$ is not possible for a spin $1/2$ particle, e.g. \cite{LL}.  Indeed, the matrix element of $\hat{Q}_E^{ij}$ can contain only the components of the angular momentum of the system, and for a spin $1/2$ particle the algebra of Pauli matrices prevents one from constructing a quadrupole moment. The presence of an external magnetic field can induce an axial deformation of the system leading to a magnetic quadrupole moment $Q^{ij}_M \sim B^i B^j$ (see e.g. \cite{Potekhin:2001xh} for the case of hydrogen atom).  

We will now argue that the chiral anomaly induces a new type of quadrupole deformation of the nucleons in the presence of a magnetic field - a ``chiral magnetic" quadrupole moment $Q^{ij} \sim \sigma^i B^j$ that involves both the components of spin and magnetic field. 
Our analysis is motivated in part by a 
recent study \cite{Eto:2011id} of the effects of chiral anomaly on the electric charge distribution inside the nucleons, and in part by the finding of the anomaly-induced quadrupole deformation of the quark-gluon plasma in a magnetic field \cite{Burnier:2011bf}.
\vskip0.2cm
{\bf 2.\,} 
In QCD the Abelian anomaly is given by~\cite{Adler:1969gk,Bell:1969ts}
\be
{\cal L}_1=\frac{N_c e^2}{12\pi^2}\,\frac{\pi^0}{f_{\pi}}\,\vec{E}\cdot\vec{B},
\label{U1}
\ee
where $\pi^0$ is the neutral pion field, $f_\pi$ is the pion decay constant, and $N_c$ is the number of colors.
The corresponding effective Hamiltonian can be obtained by taking the expectation value of (\ref{U1}) over a neutron state in the static limit:
 $$
H_1=\frac{N_ce^2}{12\pi^2}\, \times
$$
\be
\times \lim_{p'\to p}\ \int d^3x \,\vec{E}(\vec{x })\cdot\vec{B}
 \left<N(p)\left|\frac{\pi^0(\vec{x})}{f_{\pi}}\right|N(p')\right> .
\label{U2}
\ee
In this limit, the pion field is sourced by the neutron through
\be
\left(\nabla^2-m_\pi^2\right)\,\pi^0(x)=-ig_{\pi NN}\,\overline{N}\gamma_5\tau^3\,N(x) .
\label{S1}
\ee
Inserting the static solution to (\ref{S1}) in (\ref{U2}) and using the non-relativistic 
reduction for the neutron wave function yields
$$
H_1=-\frac{N_ce^2}{12\pi^2}\,\frac{g_{\pi NN}}{2M_N}\,\frac{1}{f_\pi m_\pi^2}\, \times
$$
\be
\times \ \int\,d^3x\,(\nabla^iE^j)B^j\,N^+\sigma^i\tau^3N .
\label{S2}
\ee
It is easy to see that the Hamiltonian (\ref{S2}) gives rise to the neutron's {\it chiral magnetic quadrupole moment} $Q^{ij}$:
\be
Q^{ij} =\frac{\delta H_1}{\delta \nabla^iE^j}=
-\frac{N_ce^2}{12\pi^2}\,\frac{g_{\pi NN}}{2M_N}\,\frac{1}{f_\pi m_\pi^2}\,
\,N^+\sigma^i\tau^3N\,B^j
\label{Q1}
\ee
which can be re-written as 
\be
 Q^{ij}=
-\frac{N_c\alpha}{6\pi}\,\frac{g_A}{(f_\pi m_\pi)^2}\,
\,N^+\sigma^i\tau^3N\,B^j
\label{Q2}
\ee
after the use of the Goldberger-Treiman relation. 
Note that the chiral magnetic quadrupole moment involves the correlation between different components of spin and magnetic field, and is thus different from the ``conventional" electric and magnetic quadrupole moments.

Numerically, using $g_A = 1.23$, $f_\pi = 93$ MeV and $m_\pi = 135$ MeV, we get for the chiral magnetic quadrupole moment of the neutron from (\ref{Q2}) 
\be
Q^{ij}\approx \left(1.35 \cdot 10^{-2}\,{\rm fm}^4\,\right)\,\sigma^i \,B^j .
\label{Q3}
\ee
\vskip0.2cm

{\bf 3.\,}
The chiral corrections to (\ref{Q3}) mostly contribute to the pion nucleon coupling $g_{\pi NN}$
or to the axial charge $g_A$ and renormalize them to their physical values. The non-static 
corrections to the pion field are not expected to be important for the static quadrupole
moment. The quantum chiral loops would induce higher-order  contributions to the
induced quadrupole moment; however these corrections appear to be numerically small (see below).
If a precise measurement of the chiral magnetic quadrupole moment is possible, these corrections would have to be evaluated. Such a measurement would open a possibility of a precision determination of the nucleon axial charge $g_A$. 
It is evident from the arguments presented above that the Abelian chiral anomaly affects solely the 
magnetically induced neutron quadrupole moment and not its electric charge.  Measuring this quadrupole moment is of fundamental interest and would provide a novel test of the role of the chiral anomaly in low energy QCD. Such measurements would require the presence of very strong magnetic and inhomogeneous electric fields, e.g. the field of an intense laser. 

\vskip0.2cm
{\bf 4.\,} The operator (\ref{Q2}) also appears in the low energy expansion of the amplitude of polarized Compton scattering at order $\omega^3$ ($\omega$ is the energy of the photon)\footnote{We thank M. Savage and B. Tiburzi for bringing this to our attention.}, see e.g. \cite{Holstein:2000yj,Hildebrandt:2003md,Schumacher:2005an}. Our evaluation of the quadrupole moment $Q^{ij}$ corresponds to the pion pole part of the $\sim \sigma^i B^j$ term in the spin-polarizability of the neutron. The pole contribution dominates over the higher
order chiral loops and $\Delta\pi$ contributions~\cite{L'vov:1998ez,Holstein:1999uu}.
In the case of a static external magnetic field that we consider in this note, the effect of this term can be easily understood. 
Indeed, an external magnetic field $\vec{B}$ due to the interaction with magnetic moment of the neutron will align its spin so that $\vec{\sigma} \parallel \vec{B}$. In contrast, the quadrupole moment $\sim \sigma^i B^j$ would tend to create a misalignment of the spin and magnetic field, causing the precession of the neutron's spin around the direction of magnetic field. 

\vskip0.2cm

It is now established that neutron stars and magnetars can develop very strong magnetic fields $eB \sim 10^{15}$\ G \cite{Duncan:1992hi}.
We speculate that if the neutron stars contain a $^3 P_2$ anisotropic neutron superfluid (see \cite{Page:2010aw} and references therein), the precession of the neutron spins may translate into a macroscopic precession of the angular momentum of the star. Indeed, in the presence of a strong magnetic field $\vec B$ the angular momentum $l=1$ locked to the spin $S=1$ of the Cooper pair would precess around the axis of $\vec B$. Since the Cooper pairs in a superfluid form a Bose condensate and are in the same quantum state, this precession would  
translate into a macroscopic precession of the system. This effect is similar to the coherent enhancement of microscopic parity violation in anisotropic superfluids predicted by Leggett \cite{Leggett}. 

\vskip0.2cm
 
\vskip0.2cm
We thank Martin Savage and Brian Tiburzi for bringing our attention to spin polarizabilities in Compton scattering, and to Yannis Semertzidis for the discussion of the prospects for experimental detection.
This work was supported by the U.S. Department of Energy under Contracts No.
DE-FG-88ER40388, DE-AC02-98CH10886, and DE-FG-88ER41723.

 \vfil

\end{document}